\documentstyle[prb,aps,preprint]{revtex}

\begin{document}
\draft
\title{ Larkin-Ovchinnikov-Fulde-Ferrell state in quasi-one-dimensional
superconductors }
\author{ N. Dupuis}
\address{
Laboratoire de Physique des Solides,
Universit\' e Paris-Sud,
91405 Orsay, France }
\maketitle
\begin{abstract}
The properties of a quasi-one-dimensional (quasi-1D) superconductor with
{\it
an open Fermi surface} are expected to be unusual in a magnetic field. On the
one hand, the quasi-1D structure of the Fermi surface strongly
favors the formation of a non-uniform state
(Larkin-Ovchinnikov-Fulde-Ferrell (LOFF) state) in the presence of a
magnetic
field acting on the electron spins. On the other hand, a magnetic field
acting on an open Fermi surface induces a dimensional crossover by
confining the electronic wave-functions along the chains of highest
conductivity, which
results in a divergence of the orbital critical field and in a stabilization
at low temperature of a cascade of superconducting phases separated by
first order transistions. In this paper, we study the phase diagram as a
function of the anisotropy by taking into account on the same footing the
paramagnetic and the orbital effects of the field. We discuss in details the
experimental situation in the quasi-1D organic conductors of the Bechgaard
salts family and argue that they appear as good candidates for the
observation of the LOFF state, provided that their anisotropy is large
enough. Recent experiments on the organic quasi-1D
superconductor (TMTSF)$_2$ClO$_4$ are in agreement with the results
obtained in this paper and could be interpreted as a signature of
a high-field superconducting phase.
We also point out the possibility to observe a LOFF state in some
of the recently discovered organic superconductors due to the particular
topology of their Fermi surface.
\end{abstract}
%\narrowtext

\section{Introduction}

In 1963, Larkin and Ovchinnikov and independently Fulde and Ferrell,
predicted the existence of a nonuniform superconducting state (hereafter
referred to as the LOFF state) in the presence of a magnetic field acting on
the electrons spins \cite{Larkin64}. These authors noted that the
destructive influence of Pauli paramagnetism on superconductivity can be
mitigated by pairing spin $\uparrow $ and spin $\downarrow $
electrons with a non zero total momentum whose value depends on
the magnetic field. In this way, the pairing condition, which requires that
opposite spin electrons with equal energy and a given total momentum should
be paired, can be fulfilled with improved accuracy over some
parts
of the Fermi surface. On other parts of the Fermi surface, it may then not be
possible to pair electrons at all, but the LOFF state can nonetheless be
more stable than the uniform solution. This superconducting  state occurs
only at
temperatures smaller than $T_0\simeq 0.56\, T_{c0}$ where  $T_{c0}$ is the
zero field superconducting transition temperature. The phase
transition is of first order from the LOFF state to
the ordinary uniform superconducting state and of second order to the normal
metallic phase.

Although this nonuniform state was predicted many years ago, there has been
up to now no experimental evidence of its existence \cite{Gloos93}. This can
be explained by several reasons. For an isotropic dispersion law, the
LOFF state leads only to a small increase of the zero temperature critical
field as
given by the Shandrasekhar-Clogston (or Pauli) limit \cite{Shandrasekhar62},
and its region of existence in the
$H-T$ plane, although not known exactly, is very narrow \cite{Larkin64}.
Moreover, when orbital effects of the field
are considered in a type II superconductor, the LOFF state can only exist
if the diamagnetic effect is weak enough compared to the
paramagnetic effect. The precise criterion obtained by Gruenberg and Gunther
\cite{Gruenberg66} for clean superconductors with an isotropic dispersion
law is $H_{c2}^{\rm orb}(0)/H_{P}>1.28$, where
$H_{c2}^{\rm orb}(0)$ and $H_{P}$ are the zero temperature
orbital critical field and the Pauli limited field
respectively. Finally, the LOFF state is very sensitive to impurities and is
destroyed when the elastic mean free path becomes smaller than the coherence
length \cite{Aslamazov69}.

Quasi-one-dimensional superconductors (weakly coupled chains systems
with an {\it open Fermi surface}) appear very particular with respect to the
existence of a LOFF state. The fundamental reason is that, because of the
quasi-1D structure of the Fermi surface, the partial compensation of the
Pauli pair breaking (PPB) effect by a spatial modulation of the order
parameter
is much more efficient than in a system with an isotropic dispersion law
\cite{Aoi74,Dupuis93a,Dupuis93}. Moreover, it has been shown recently that
the magnetic field induces a dimensional crossover which makes the orbital
critical field $H_{c2}^{\rm orb}$ diverge \cite{Lebed86,Dupuis93a,Dupuis93},
thus increasing
the relative strength of the PPB effect compared to the orbital
effect. Noting also that quasi-1D systems such as can be found
experimentally in the organic conductors of the Bechgaard salts family can
be made very clean, strongly anisotropic superconductors should be very good
candidates for the observation of a LOFF state.

The effect of a magnetic field on the phase diagram of
a quasi-one-dimensional superconductor has recently received a lot of
attention \cite{Lebed86,Dupuis93a,Dupuis93,ND}. It was shown that for
$T_{c0}\ll
t_z$ a high magnetic field stabilizes at low temperature a cascade of
superconducting phases separated by first order transitions, which ends in a
strong reentrance of the superconducting phase (the magnetic field is
parallel to the $y$ axis; $t_z$ is the coupling in the $z$
direction between the chains parallel to the $x$ axis). The existence of
this cascade of superconducting phases in high magnetic field is a
consequence of the two properties of a quasi-one-dimensional
superconductor noted above: The magnetic-field-induced dimensional
crossover which freezes the orbital mechanism of destruction of the
superconductivity and the efficiency of the LOFF state in compensating
the PPB effect \cite{Lebed86,Dupuis93a,Dupuis93}. In the reentrant phase,
the dimensional
crossover is almost complete: The system exhibits a quasi-2D behavior
and the critical temperature is mainly determined by the PPB effect.
For small enough $t_z$ ($t_z \sim T_{c0}$), the cascade of phase transition
disappears and the reentrant phase
appears more as a slow decrease of the critical temperature than as
a real reentrance of the superconducting phase.
The critical temperature $T_c$ decreases as $\sim 1/H$, a consequence of the
existence of a LOFF state, leading to an upward curvature of the critical
field $H_{c2}(T)$ \cite{Dupuis93}.

In this paper, we determine the transition line
$T_c(H)$ (or $H_{c2}(T)$) as a function of the parameter $t_z/T_{c0}$ in the
presence of both orbital and PPB effects. The present work is an extension
of the work of Dupuis and Montambaux (hereafter DM) \cite{Dupuis93} with
special attention devoted  to the appearance of a LOFF state and
to the crossover between the low field regime and the high field regime (or
quantum regime in the terms used in Ref.\cite{Dupuis93}).

In the next section, we determine the transition line
in the absence of orbital effect of the field. At low field, $T_c-T_{c0}
\sim (\mu _BH)^2/T_{c0}$ which leads to a downward curvature of the upper
critical field. We show
that below $T_0\simeq 0.56\, T_{c0}$, the LOFF state is more stable than the
uniform superconducting state. For $\mu _BH\gg T$, the critical temperature
between the LOFF state and the normal state varies
as $1/H$ leading to an upward curvature of the critical field $H_{c2}(T)$.
The effect of disorder is discussed.
In section \ref{OPE}, we study the effect of a small coupling $t_z\sim
T_{c0}$ between chains on the phase diagram obtained in section \ref{PP}.
For a large anisotropy (i.e., $t_z/T_{c0}\sim 1$), the phase diagram
obtained in section \ref{PP} is only slightly modified by the orbital
effects. At low field, the critical temperature $T_c$ is now dominated by
the orbital effects of the field and decreases linearly with $H$. However,
the upward curvature of $H_{c2}(T)$ at $\mu _BH\gg T$, which results from
the existence of a LOFF state, subsists when the quantum effects of the
field are fully taken into account in the calculation of $T_c$. For
a smaller anisotropy (i.e., $t_z/T_{c0}> 2$), the phase diagram becomes
more complicated due to the stabilization at low temperature of the cascade
of superconducting phases studied by DM. The interplay between this cascade,
which is induced by the orbital effects of the field, and the appearance of
the LOFF state is studied in details as a function of the anisotropy
$t_z/T_{c0}$. In section \ref{BS}, we discuss the experimental situation in
the quasi-1D superconductors of the Bechgaard salts family and argue that
they appear as good candidates for the
obvervation of a LOFF state, provided that their anisotropy is large
enough. Recent experimental results obtained by Lee {\it et al.}
\cite{Lee94} with the organic quasi-1D superconductor (TMTSF)$_2$ClO$_4$
(TMTSF=tetramethyltetraselenafulvalene) are discussed.
In the conclusion, we point out the possibility to observe a LOFF state in
some of the recently discovered quasi-2D organic superconductors due to the
particular topology of their Fermi surface.

\section{PAULI PARAMAGNETISM}
\label{PP}

In this section, we consider a strongly anisotropic superconductor
subject to a magnetic field acting on the electron spins.
The open Fermi surface is described by the dispersion law
($\hbar =k_B=1$ in the following and the Fermi energy is chosen as the
origin of the energies)
\begin{equation}
\epsilon _{\bf k}^\alpha =v(\alpha k_x-k_F) +t_z \cos (k_zc)  \,,
\label{LDL}
\end{equation}
where $v$ is the Fermi velocity for the motion along the chains ($x$ axis)
and $c$ the distance between chains. $\alpha ={\rm sgn}(k_x)=+/-$ labels
the right/left sheet of the Fermi surface. We do not consider explicitely
the $y$ direction parallel to the magnetic field which does
not play any role for a linearized dispersion law.

At high temperature (or low magnetic field), the order parameter $\Delta $
is uniform. Its value is obtained from the self-consistency equation
\begin{equation}
{1 \over \lambda }={T \over S} \sum _{{\bf k},\omega _n}
{1 \over {{\epsilon _{\bf k}^\alpha }^2-(i\omega _n+h)^2+\Delta ^2}} \,,
\label{eqgap}
\end{equation}
where $\lambda >0$ denotes the BCS attractive interaction and $h=\mu _BH$ is
the Zeeman energy (the $g$ factor is assumed to be equal to 2). $\omega
_n=\pi T(2n+1)$ is a Matsubara frequency and $S$ is the area of the system.
The difference $F(T,H)$ between the free energies of the superconducting
state and of the normal state can be obtained from \cite{Fetter}
\begin{equation}
F(T,H)=\int _0^\Delta {{dg} \over {d\Delta '}} {\Delta '}^2 d\Delta ' \,,
\end{equation}
where the function $g(\Delta )=1/\lambda $ is defined by (\ref{eqgap}).
Expanding the self-consistency equation in powers of $\Delta $, we obtain the
Ginzburg-Landau (GL) expansion of the free energy
\begin{equation}
F(T,H)=A\Delta ^2+{B \over 2}\Delta ^4 +{C \over 3} \Delta ^6 \,,
\end{equation}
where
\begin{eqnarray}
A &=& \lambda ^{-1}-\chi (0) \,, \nonumber \\
B &=& {{N(0)} \over 2} {1 \over {(2\pi T)^2}}
{\rm Re }\, \zeta \Biggl (3,{1 \over 2}+{h \over {2i\pi T}} \Biggr )
\,, \nonumber \\
C &=& -{{3N(0)} \over 8} {1 \over {(2\pi T)^4}}
{\rm Re }\, \zeta \Biggl (5,{1 \over 2}+{h \over {2i\pi T}} \Biggr ) \,,
\label{GLexp}
\end{eqnarray}
where $\zeta (a,z)=\sum
_{n=0}^\infty (n+z)^{-a}$ is the generalized zeta function. $N(0)$ is the
density of states per spin at the Fermi level.
$\chi (0)$ is the Cooper pair susceptibilty
\begin{eqnarray}
\chi (q) &=& {T \over S} \sum _{{\bf k},\omega _n}
{1 \over { (i\omega _n-\epsilon _{\bf k}^\alpha -h)
(-i\omega _n-\epsilon _{q-{\bf k}}^{-\alpha }+h) }}
\nonumber \\
&=& N(0) \Biggl \lbrack \ln \Biggl ( {{2\gamma \Omega } \over {\pi
T}} \Biggr ) + \Psi \Biggl ( {1 \over 2} \Biggr )-{1 \over 2} \sum
_\alpha {\rm Re}\, \Psi \Biggl ( {1 \over 2} +{{\alpha vq+2h} \over {4i\pi
T}} \Biggr ) \Biggr \rbrack
\label{chi}
\end{eqnarray}
evaluated at zero total momentum and $\Psi $ is the digamma function.
Here $q$ is the momentum along the chains. $\gamma $ is the exponential of
the Euler constant and $\Omega $ is the cutoff energy for the attractive
interaction.
The GL expansion (\ref{GLexp}) agrees with the one obtained
by Maki and Tsuneto in the case of an isotropic system with the dispersion
law $\epsilon _{\bf k}=k^2/2m$ \cite{Maki64}. As long as the Cooper pairs
are formed with states of opposite momenta, the shape of the Fermi surface
does not play any role. This appears clearly when we make the usual
replacement $S^{-1}\sum _{\bf k} \rightarrow N(0)\int d\epsilon $ in the
self-consistency equation (\ref{eqgap}). The critical temperature is
determined by $A=\lambda ^{-1}-\chi (0)=0$. For low field $h\ll T$,
using $\lambda ^{-1}=N(0)\ln (2\gamma \Omega /\pi T_{c0})$, we
obtain a downward curvature of the transition line (or equivalently of the
critical field $H_{c2}(T)$):
\begin{equation}
T_{c0}-T_c\simeq {{7 \zeta (3)} \over {4\pi ^2}} {{h^2} \over {T_{c0}}} \,,
\end{equation}
where $\zeta (3)\simeq 1.20$.
As pointed out by Maki and Tsuneto \cite{Maki64,Sarma63}, the transition
between the normal
state and the uniform superconducting state becomes of first order when the
coefficient  of the quartic term in the GL expansion changes sign. This
corresponds to the point $(h_0,T_0)$ of the transition line determined by
\begin{equation}
{\rm Re}\, \zeta \Biggl (3,{1 \over 2}+{h_0 \over {2i\pi T_0}} \Biggr )=0
\; {\rm  and }\; A=0 \,.
\label{pt0}
\end{equation}
The first equality in (\ref{pt0}) leads to $h_0/2\pi T_0\simeq 0.304092$.
{}From $A=0$ and $\lambda ^{-1}=N(0)\ln (2\gamma \Omega /\pi T_{c0})$, we then
deduce $T_0\simeq 0.56\, T_{c0}$.

Up to now, we have ignored the possibilty to observe a nonuniform
superconducting state. It is therefore necessary to consider the more
general equation
\begin{equation}
{1 \over \lambda }=\chi (q) \,,
\end{equation}
where $\chi (q)$ is the Cooper pair susceptibility defined in (\ref{chi}).
In principle, one should also consider the possibilty of a
nonuniform superconducting state with a finite momentum of the Cooper pair
along the $z$ direction. We have verified that in our model such a state is
never stable.
The wave vector of the modulation of the order parameter at the transition is
determined by the maximum of the susceptibility, which can be obtained
from the derivatives
\begin{eqnarray}
\chi '(q)&=&-N(0){v \over {8\pi T}} \sum _\alpha {\rm Re}\, {\alpha \over i}
\Psi ' \Biggl ( {1 \over 2} +{{\alpha vq+2h} \over {4i\pi T}} \Biggr ) \,,
\label{chip} \\
\chi ''(q) &=& -N(0) \Biggl ( {v \over {4\pi T}} \Biggr )^2
\sum _\alpha {\rm Re}\, \zeta  \Biggl (3,{1 \over 2}+
{{\alpha vq+2h} \over {4i\pi T}}  \Biggr ) \,,
\end{eqnarray}
where $\Psi '$ and $\Psi ''$ are the first and second derivatives of the
digamma
function. The last equation was obtained using $\Psi ''(z)=-2\zeta (3,z)$.
It can be seen from Eq.(\ref{chip}) that $\chi '(0)=0$ independently
of the value of the field. For $h/T<h_0/T_0$, $\chi ''(0)<0$ so that $q=0$
corresponds to a maximum of the susceptibility. For $h/T>h_0/T_0$, $\chi
''(0)>0$ showing that $q=0$ corresponds to a local minimum of the
susceptibilty. The maximum of $\chi (q)$ is reached for a finite value of
the total momentum. Thus, the temperature $T_0$ below which the LOFF state
is more stable than the uniform superconducting state corresponds exactly to
the temperature below which we showed that the transition between the normal
state and the uniform superconducting state would have become of first order
in the absence of the LOFF state. An anologous result has been obtained by
Dieterich and Fulde in their study of the magnetic field dependence of the
Peierls  instability in one-dimensional conductors, a problem which bears
some similarities with the one considered in this section
\cite{Dieterich73}.
For $h/T$ slightly below $h_0/T_0$, we find using Eq.(\ref{chip}) that the
maximum of the susceptibilty is obtained for
\begin{equation}
q^2=6 \Biggl ( {{4\pi T} \over v} \Biggr )^2
{ {{\rm Re} \, \Psi '' \Bigl ( {1 \over 2}+{h \over {2i\pi T}} \Bigr ) }
\over
{{\rm Re} \, \Psi ^{(4)} \Bigl ( {1 \over 2}+{h \over {2i\pi T}} \Bigr ) }
} \,,
\end{equation}
where $\Psi ^{(4)}$ is the fourth derivative of the digamma function.
For large field $h\gg T$, Eq.(\ref{chi}) shows that the maximum of the
susceptibilty
should be reached for $q\simeq \pm 2h/v$. This leads to the critical
temperature
\begin{equation}
T_c \simeq {{\pi T_{c0}^2} \over {4\gamma h}} \,,
\end{equation}
a result which was previously obtained by DM. Thus, at low
temperature, the variation of $T_c$ as $1/H$, which is a consequence of the
existence of the LOFF state, leads to a divergence and an upward curvature
of the critical field $H_{c2}(T)$.
The susceptibilty $\chi (q)$ as a function of $q$ is shown
in Fig.\ref{Fig1} for different values of the magnetic field. The transition
line $T_c$ and the wave vector $q$ of the order parameter are shown in
Fig.\ref{Fig2}.

The divergence of the critical field is of course not physical. At low
temperature, the effect of disorder will become more and more important and
will lead to a finite critical field. Following the
standard treatment \cite{Gorkov60}, impurity scattering is taken into
account by including self-energy and vertex corrections in the Cooper pair
susceptibity. Using the results of Ref.\cite{Dupuis93}, we find that the
critical temperature in presence of disorder $T_c^{\rm dis}$ is given by
\begin{equation}
{{T_c^{\rm dis}-T_c} \over {T_c}} \simeq -{\pi \over {32T_c \tau }}
+ {{3\pi } \over {32T_c \tau }} {{T_c^2} \over {h^2}}
\end{equation}
for $\vert T_c^{\rm dis}-T_c \vert \ll T_c \ll h$. Thus, the disorder
becomes
important at low temperature when $T_c \sim \pi /32\tau $. In Bechgaard
salts where $1/\tau $ can be of the order of 100 mK, the disorder will be
inefficient down to very low temperature  so that the upward curvature of
the upper $H_{c2}(T)$ will persist in a very broad range of temperature.

In the present model, the strong stability of the LOFF state strongly relies
on the use of a linearized dispersion law. It is therefore necessary to
verify that a finite curvature of the dispersion law at the Fermi level does
not modify significantly the preceding results. Instead of the linearized
dispersion law given by Eq.(\ref{LDL}), we consider the following
tight-binding model:
\begin{equation}
\epsilon _{\bf k}= t_x\cos (k_xa)+t_y\cos (k_yb)+t_z\cos (k_zc)-\mu \,,
\end{equation}
where $\mu $ is the Fermi energy, $a$, $b$ and $c$ the lattice parameters.
The transfer integrals $t_x$, $t_y$ and $t_z$ verify the condition $t_y,t_z
\ll
t_x$ which ensures that the Fermi surface is open for a sufficient filling
($\mu \sim t_x$).
If we expand $\epsilon
_{\bf k}$ around $\pm k_F$ defined by $\mu =t_x \cos (k_Fa)$, we obtain
\begin{equation}
\epsilon _{\bf k} \simeq v(\vert k_x \vert -k_F) +{1 \over 2}
(\vert k_x \vert -k_F)^2a^2t_x\cos (k_Fa) +t_y\cos (k_yb)+t_z\cos (k_zc) \,,
\end{equation}
where $v=at_x\sin (k_Fa)$ is the Fermi velocity along the chains direction.
Because of the curvature of the dispersion law around the Fermi level, it is
not possible to find a particular value of $q$ allowing us to
fulfill the pairing condition
$\epsilon _{\bf k}+h=\epsilon _{\bf q-k}-h$ for one half of the phase space.
However, it will be possible to neglect
the curvature of the dispersion law if
\begin{equation}
{1 \over 2}q^2a^2t_x\cos (k_Fa) \ll T \,,
\end{equation}
where $q\sim \pm 2h/v$ is the total momentum of the Cooper pair in the LOFF
state. Using $k_Fa\sim 1$ and $t_x\sim vk_F$, this inequality can be
rewritten as
\begin{equation}
h^2 \ll Tt_x \,.
\end{equation}
For $t_x/2 \sim 3000$ K and using $h=\mu_BH=0.67\times H$ K, we obtain the
condition: $H\ll 30 $ T at $T=100$ mK. Thus, in the magnetic field and
temperature ranges which can be
experimentally reached, the use of a linearized dispersion law is justified.

\section{ORBITAL AND PAULI EFFECTS}
\label{OPE}

In this section, we study how the orbital effects of the magnetic field
modify the phase diagram obtained in the preceding section.
In the gauge ${\bf A}(0,0,-Hx)$, the order parameter is determined by the
integral equation \cite{Lebed86,Dupuis93a,Dupuis93}
\begin{equation}
\lambda ^{-1} \Delta (x,q_z)= \int _{\vert x-x' \vert >d } dx' K(x,x',q_z)
\Delta (x',q_z) \,,
\label{eqINT}
\end{equation}
\begin{eqnarray}
K(x,x',q_z) &=& {{N(0)\pi T} \over v}
{{\cos \lbrack 2\mu _BH(x-x')/v \rbrack } \over {\sinh \lbrack \vert x-x'
\vert 2\pi T/v \rbrack  }}
\nonumber \\ & & \times
J_0 \Biggl ( {{4t_z} \over \omega _c} \sin \Biggl \lbrack {G \over 2}
(x-x') \Biggr \rbrack \sin \Biggl \lbrack q_z {c \over 2} -{G \over 2}
(x+x') \Biggr \rbrack \Biggr ) \,,     \label{kernel}
\end{eqnarray}
where $K$ takes into account both the PPB and orbital effects. $J_0$ is the
zeroth order Bessel function, $G=-eHc$ and $\omega _c=Gv$.
The cut-off $d$ is related to the energy $\Omega $. Taking advantage of the
conservation of the transverse momenta in the chosen gauge, we have
introduced the Fourier
transform $\Delta (x,q_z)$ of the order parameter with respect to $z$.
$q_z$ only shifts the origin of the $x$
axis and can therefore be set equal to zero when determining the critical
temperature. Without any loss of generality, the solution of the integral
equation (\ref{eqINT}) can be written as a Bloch function \cite{Dupuis93}
\begin{equation}
\Delta _Q(x)=e^{iQx} \tilde \Delta _Q(x) \,,
\end{equation}
where $\tilde \Delta _Q(x)$ has the periodicity $\pi /G$ and the magnetic
Bloch wave vector $Q$ is restricted to $\rbrack -G,G \rbrack $. Each phase
is characterized by this vector $Q$ which plays the role of a
pseudo-momentum
for the Cooper pairs in the magnetic field. The kernel $K(x,x')$ takes into
account all the quantum effects of the field. In Sec.\ref{OPELA} we
will compare
the exact mean-field results obtained with $K(x,x')$ with those obtained in
the eikonal (or semiclassical phase integral) approximation where the
quantum effects of the field are completely neglected. In this
approximation, the kernel becomes \cite{Dupuis93}
\begin{eqnarray}
K^{\rm (eik)}(x,x',q_z) &=& {{N(0)\pi T} \over v}
{{\cos \lbrack 2\mu _BH(x-x')/v \rbrack } \over {\sinh \lbrack \vert x-x'
\vert 2\pi T/v \rbrack  }} \nonumber \\ & & \times
J_0 \Biggl ( {{2t_z} \over v}
(x-x') \sin \Biggl \lbrack q_z {c \over 2} -{G \over 2}
(x+x') \Biggr \rbrack \Biggr ) \,.     \label{kerneleik}
\end{eqnarray}

If $t_z=0$, the orbital effects vanish for a field parallel to the $y$
direction and the phase diagram is then shown in Fig.\ref{Fig2}. In the
following, we study the orbital effects of the field for increasing
coupling between chains.

\subsection{Large anisotropy }
\label{OPELA}

We first consider the case of a small coupling $t_z/T_{c0}=1.33$.
For each value of the field, we determine numerically
from Eq.(\ref{eqINT}) the vector $Q$ which maximizes the critical
temperature and the corresponding $T_c$.

In a first step, we neglect the PPB effect. The results are shown in
Figs.\ref{Fig3} and \ref{Fig4}.
In the eikonal approximation (Fig.\ref{Fig3}), where quantum effects of
the field are not taken into account, we
recover the standard results for a system of weakly coupled superconducting
chains (or planes) \cite{Bulaevskii90}. Close to $T_{c0}$, the critical
temperature decreases linearly with $H$. In this regime, the coherence
length $\xi _z(T)$ is much larger than the spacing $c$ between chains and
the superconducting state is a triangular Abrikosov vortex lattice. At
lower temperature, the coherence length becomes of the order of the spacing
between chains so that vortices can fit between chains, thus quenching the
orbital mechanism of destruction of the superconductivity and leading to a
divergence of the critical field. The superconducting state
is then a triangular Josephson vortex lattice with a periodicity
in the transverse direction equal to $2c$. The crossover
between these two regimes is sometimes referred to as a
(temperature-induced) dimensional crossover \cite{Klemm75}. It should be
noted here
that this dimensional crossover is different from the magnetic-field-induced
dimensional crossover which results from the magnetic-field-induced
localization of the wave functions and which is not taken into account in
the eikonal approximation \cite{Dupuis93}. The values of $Q$ corresponding
to the highest $T_c$ are shown in Fig.\ref{Fig3}b. At low field,
all the values of $Q$ lead to
the same critical temperature. As pointed out by DM, this degeneracy allows
one to construct the Abrikosov vortex lattice by taking a linear combination
of the function $\Delta _Q(x,q_z)$. In Fig.\ref{Fig3}b, the degeneracy of
$T_c$ with respect to $Q$ is shown symbolically by a shaded triangle.
At higher field, when the superconducting state
becomes a Josephson vortex lattice, the degeneracy is lifted in favor of
$Q=0$. It is worth pointing out that these results obtained in the eikonal
approximation can also be obtained in the
Lawrence-Doniach model \cite{LD70} where the critical temperature is
obtained from (restoring the $q_z$ dependence) \cite{Dupuis93}
\begin{equation}
-v^2 {{\partial ^2\Delta } \over {\partial x^2}}
+t_z^2 \Bigl \lbrack 1-\cos (q_zc-2Gx) \Bigr \rbrack \Delta =
{{16 \pi ^2} \over {7\zeta (3)}} T_{c0}^2
\Biggl ( 1- {{T_c} \over {T_{c0}}} \Biggr ) \,.
\end{equation}
The results obtained in the exact calculation are similar to those obtained
in the eikonal approximation, except for the reentrance which occurs at high
field ($\omega _c \gg t_z$) as a consequence of the magnetic-field-induced
dimensional crossover \cite{Dupuis93} (Fig.\ref{Fig4}).

We now consider both the PPB and orbital effects. In both descriptions
(exact and eikonal), we obtain a linear behavior at low field showing that
the critical temperature is limited by the orbital effect. The value of $Q$
is degenerate in this field range. At higher field, the degeneracy is
lifted in favor of $Q=0$ when the periodicity of the vortex lattice becomes
of the order of $2c$. This temperature-induced dimensional crossover from
the Abrikosov vortex
lattice towards the Josephson vortex lattice is accompanied by a weak upward
curvature of $H_{c2}(T)$. At lower temperature, the two descriptions
differ considerably. In the exact description (Fig.\ref{Fig5}), the orbital
effect appears
very weak and the phase diagram in this field range is similar to the one
obtained by considering only the PPB effect (Fig.\ref{Fig2}). We observe
a transition to a LOFF state characterized by a finite value of $Q$ which
means an additional spatial modulation for the Josephson vortex lattice.
For very high field, we find $Q\simeq 2h/v$. The
transition line shows a pronounced upward curvature which is a consequence
of the existence of the LOFF state. In the eikonal description
(Fig.\ref{Fig6}),
the orbital effect modifies in an important way the phase diagram shown in
Fig.\ref{Fig2}. The divergence of the critical field $H_{c2}$ is suppressed
\cite{Bulaevskii73}
and the region of stability of the LOFF state is very narrow. The upward
curvature of the transition line is now restricted to very low
temperatures.

Thus, the pronounced upward curvature of $H_{c2}(T)$ found in the preceding
section, which was a consequence of the existence of the LOFF state,
persists only if the quantum effects of the field are fully taken into
account. In the following, we shall not consider the eikonal
approximation any more.

\subsection{Smaller anisotropy }

For larger values of the coupling between chains $t_z/T_{c0}=2.67$ and 2.93,
the phase diagrams are shown in Fig.\ref{Fig7} to Fig.\ref{Fig10}.
The low field regime, where the value of
$Q$ is degenerate, is now followed by a phase $Q=G$, which is itself
followed
by a phase $Q=0$. This is the cascade of superconducting phases which has
been studied by DM. This cascade appears
between the low field regime where the superconducting state is an
Abrikosov vortex lattice and the very high field regime where the
superconducting state is a Josephson vortex lattice. The transition to the
LOFF state appears in the last phase $Q=0$: The GL regime and the cascade of
phases are dominated by the orbital effects of the field.
Fig.\ref{Fig7} to Fig.\ref{Fig10} show that the shape of the transition line
is very sensitive to the value of $t_z/T_{c0}$.

If we further increase the coupling between chains (Fig.\ref{Fig11} to
Fig.\ref{Fig14}), the number of phases
in the cascade increases. The transition to the LOFF state appears before the
reentrant phase. In Fig.\ref{Fig12}, the transition corresponds to a shift
of $Q$ within a phase $Q=G$.

For $t_z/T_{c0}=6.67$ (Fig.\ref{Fig15} and Fig.\ref{Fig16}), the cascade of
phase transitions appears at lower
temperature. The transition to the LOFF state occurs in the beginning of
the cascade. At low temperature, we thus observe an alternance of phases
$Q=2h/v$ and $Q=G-2h/v$. In Fig.\ref{Fig17}, we have shown the eigenvalue
$\lambda _Q$ of the kernel $K(x,x')$ associated with the eigenfunction
$\Delta _Q(x)$ for different values of the magnetic field. The parameters
used in this figure are the same as the ones of Fig.\ref{Fig16}.

\section{BECHGAARD SALTS}
\label{BS}

In this section, we discuss the experimental situation in the Bechgaard
salts. We concentrate on (TMTSF)$_2$ClO$_4$ which appears as the
most promising material with respect to the effects discussed in this paper.
Several comments are in order here. First we should wonder whether the model
we have used is adequate to describe the superconductivity in the Bechgaard
salts. Many experimental results seem to show that the
intrachain interactions are repulsive and indicate that the
superconductivity is not of conventional type \cite{Jerome82,CB91a}. For
example, NMR measurements
obtained by Takigawa \cite{Takigawa87} are clearly not compatible with
isotropic s-wave pairing in zero field. However, DM argued that a model
based on local attractive interactions (which would lead to
isotropic s-wave pairing in zero field) should remain valid from a
qualitative point of view. The reason is that the unusual behavior of a
quasi-1D superconductor in a high magnetic field is due only to the
magnetic-field-induced dimensional crossover (i.e., the localization of
the one-particle wave-functions along the chains of highest conductivity)
and does not rely
on a particular model of superconductivity \cite{local}. Second, one should
wonder whether a BCS mean-field analysis can be justified in a system of
weakly
coupled chains. Such an analysis requires well defined quasi-particles in the
normal state and in particular a coherent transverse (in the $y$ and $z$
directions) electronic motion. From a theoretical point of view, this
problem has recently attracted a lot of attention. Many authors
\cite{CB91} have given some arguments in favor of a Fermi liquid behavior
at low enough temperature in a system of weakly
coupled chains \cite{CB91b} while the opposite point of view has been
adopted by Anderson and collaborators
\cite{Anderson91}. In (TMTSF)$_2$ClO$_4$,
the NMR relaxation rate \cite{Wzietek93}
shows that the behavior of the system changes drastically when the
temperature decreases below about 10-30 K. This result, together with the
absence of a correlation gap as can be seen from resistivity measurement,
strongly suggests that this compound undergoes a single particle
dimensionality crossover at a temperature $T_{x^1}\sim 10$ K, below which
the tranverse electronic motion becomes coherent. Other
convincing experiments are those concerned with
the angular Lebed' resonances \cite{Lebed86a}. The origin of these
resonances, which occur
in various physical quantities (thermodynamics or transport) when the field
is titled in the $(y,z)$ plane, can be simply understood from  a
semiclassical argument. The semiclassical electronic trajectory is
of the form $y=b(t_y/\omega _{cy})\cos (\omega _{cy}y/v)$ and
$z=c(t_z/\omega _{cz})\cos (\omega _{cz}z/v)$, where $\omega _{cy}=-eHc\cos
(\theta )$ and $\omega _{cz}=-eHb\sin (\theta )$. Here $b$ is the spacing
between chains in the $y$ direction and $\theta $ is the angle between ${\bf
H}$ and the $z$ axis.
For certain orientations $\theta $ of the field (``magic"
angles), the two magnetic frequencies $\omega _{cy}$ and $\omega _{cz}$ are
comensurate,
$\omega _{cz}/\omega _{cy}=p/q$ ($p,q$ integer), leading to a periodic
electronic motion which results in the Lebed' resonances. Clearly,
this analysis based on the consideration of the semiclassical orbits is
meaningful only if the
electronic motion is coherent in both the $y$ and $z$ directions. From
our point of view,
the absence of coherent transverse electronic motion in the Bechgaard salts
would therefore be very difficult to reconcile with the existence of these
angular oscillations \cite{Strong94}.

Since the shape of the transition line  very strongly depends
on the value of the ratio $t_z/T_{c0}$, one should wonder
what the value of this parameter is in the Bechgaard salts.
There are two opinions in
the literature concerning the values of the transfer integrals $t_b=t_y/2$
and $t_c=t_z/2$ \cite{tc}. According to many authors, $t_b=200-300$ K and
$t_c=5-10$ K, as obtained from band calculation \cite{Ducasse85}. These
values seem to be supported by recent measurements of a new type of angular
oscillations of the conductivity \cite{Danner94}. The second opinion is
that the values of $t_b$ and $t_c$ are strongly reduced with respect to their
bare values, due to 1D fluctuations \cite{CB91a,CB91}. From a
renormalization
group calculation and using experimental results of the NMR relaxation rate,
Bourbonnais {\it et al.} estimated $t_b\simeq 30$ K \cite{Wzietek93}, which
leads to $t_c\simeq 0.5-1.5$ K. While the first point of view yields values
of the transfer integrals which would make the observation of the LOFF state
difficult or even impossible (see the numerical calculations in
Sec.\ref{OPE} and Ref.\cite{Dupuis93}), the second point of view makes the
compound (TMTSF)$_2$ClO$_4$ a very good candidate for the effects discussed
in this paper.

Lee {\it et al.} have recently investigated the superconducting transition
in (TMTSF)$_2$ClO$_4$ from resistive measurements performed between 1.2 K
and 60 mK and up to 7 Tesla \cite{Lee94}. The magnetic field was oriented
along the b' axis, since this corresponds to the orientation for which
quantum effects of the field are expected \cite{Dupuis93}. Although they do
not give a definite answer for the existence of high-field
superconductivity in (TMTSF)$_2$ClO$_4$, these results might be interpreted
as the signature of a high-field superconducting phase (see
Ref.\cite{Lee94} for a detailed analysis of the experimental results). Such
an interpretation would imply an anisotropy $t_z/T_{c0}\sim 3.5-4$
which leads to $t_c \sim 2$ K. For this
value of the anisotropy, the GL regime and the reentrant phase are separated
by a few superconducting phases separated by first order transitions
(see Figs. \ref{Fig12} and \ref{Fig14}).

\section{CONCLUSION}

We have presented a detailed analysis of the interplay of the Pauli
paramagnetism and the orbital effects of the field in a quasi-1D
superconductor with an open Fermi surface. We have calculated the
transition line $(H,T_c)$ as a function of the anisotropy $t_z/T_{c0}$.
As a result of their quasi-1D
Fermi surface, quasi-1D superconductors appear as very good candidates
for the observation of a LOFF state, provided that their anisotropy is
large enough. We have shown that when the interchain coupling is
sufficiently weak, the transition between the LOFF state and the normal
state is characterized by an upward curvature of the critical field
$H_{c2}(T)$. We have also argued that the organic superconductor
(TMTSF)$_2$ClO$_4$
is a good candidate for the observation of a LOFF state. The experimental
results of Lee {\it et al.} \cite{Lee94} on this compound are in
agreement with the results
obtained in this paper and could be interpreted as a signature of high-field
superconductivty. Since the zero field critical temperature
$T_{c0}$ decreases with pressure \cite{Kang93}, it should be possible to
study the evolution of the resistivity curves $R(H,T)$ measured  by Lee {\it
et al.} as a function of the anisotropy $t_z/T_{c0}$. A disappearence of the
high-field ``anomalies" of the resistivity when pressure is increased would
support the existence of a high-field superconducting phase at ambiant
pressure.
Although we only considered in this paper the case of quasi-1D
systems, the results can easily be extended to the case of quasi-2D
conductors. DM argued that similar quantum effects of the field should be
present in a weakly coupled planes system in a magnetic field perpendicular
to the low conductivity axis, leading to a cascade of superconducting
phases similar to the one predicted for quasi-1D systems.
They also pointed out that for a dispersion law which is isotropic in the
most
conducting plane (for example $\epsilon _{{\bf k}_\parallel }=k_\parallel
^2/2m_\parallel $ where ${\bf k}_\parallel $ and $m_\parallel $ are the wave
vectors and the effective mass in the most conducting plane), the LOFF state
will not compensate significantly the PPB effect so that its region of
stability in the $H-T$ plane will be very narrow or even non-existent. This
means that both the
LOFF state and the cascade will be very difficult to observe in this case.
However, some of the recently discovered
quasi-2D organic superconductors, like for example the salts
of the BEDT-TTF family, present some
flat parts on their Fermi surface \cite{Mori94}. While such a
topology of the Fermi surface is usually expected to favor the
formation of an antiferromagnetic state, it could also
increase the efficiency of the LOFF state in compensating the PPB effect.
Experimental results indicate that the critical field parallel to the most
conducting planes is closed to the Pauli limit \cite{Mori94}. Thus, the
existence of a LOFF state in quasi-2D materials cannot be a priori excluded.

\section*{ACKNOWLEDGMENTS }

The author would like to thank G. Montambaux for useful comments and a
critical reading of the manuscript, and K. Behnia for useful discussions.
The Laboratoire de Physique des Solides is Unit\'e Associ\'ee au CNRS.

\begin{figure}
\caption{Cooper pair susceptibility $\chi (q)$ on the transition line
$(H,T_c)$ as a function of $q/G$ for different values of the magnetic field.
$G=-eHc$ is the magnetic wave vector introduced in section III. }
\label{Fig1}
\end{figure}

\begin{figure}
\caption{a) Transition line $T_c$ of a quasi-1D superconductor in presence
of a magnetic field acting on the electrons spins. b) Wave vector $q$ of the
modulation of the order parameter vs magnetic field. The dashed line
corresponds to $q=2h/v$. }
\label{Fig2}
\end{figure}

\begin{figure}
\caption{a) Transition line $T_c$ for $t_z/T_{c0}=1.33$ in the eikonal
approximation and in the absence of PPB effect. On the top of the figure,
we have written (Eik, no PPB) in order to remind that the calculation was
done in the eikonal approximation without considering the Pauli pair
breaking
effect. b) Magnetic Bloch wave vector $Q$ vs magnetic field. The degeneracy
of $T_c$ with respect to $Q$ at low field is shown symbolically by a
shaded triangle. The three dashed lines correspond to $Q=2h/v,G-2h/v,G$. }
\label{Fig3}
\end{figure}

\begin{figure}
\caption{a) Transition line $T_c$ for $t_z/T_{c0}=1.33$ in the absence of
PPB effect. b) Magnetic Bloch wavevector $Q$ vs magnetic field. }
\label{Fig4}
\end{figure}

\begin{figure}
\caption{a) Transition line $T_c$ for $t_z/T_{c0}=1.33$ in the presence of
both PPB and orbital effects. b) Magnetic Bloch wavevector $Q$ vs magnetic
field. }
\label{Fig5}
\end{figure}

\begin{figure}
\caption{a) Transition line $T_c$ for $t_z/T_{c0}=1.33$  in the eikonal
approximation in the presence of both PPB and orbital effects. b) Magnetic
Bloch wavevector $Q$ vs magnetic field. }
\label{Fig6}
\end{figure}

\begin{figure}
\caption{a) Transition line $T_c$ for $t_z/T_{c0}=2.67$
in the absence of PPB effect. b) Magnetic
Bloch wavevector $Q$ vs magnetic field. }
\label{Fig7}
\end{figure}

\begin{figure}
\caption{a) Transition line $T_c$ for $t_z/T_{c0}=2.67$
in the presence of both PPB and orbital effects. b) Magnetic
Bloch wavevector $Q$ vs magnetic field. }
\label{Fig8}
\end{figure}

\begin{figure}
\caption{a) Transition line $T_c$ for $t_z/T_{c0}=2.93$
in the absence of PPB effect. b) Magnetic
Bloch wavevector $Q$ vs magnetic field. }
\label{Fig9}
\end{figure}

\begin{figure}
\caption{a) Transition line $T_c$ for $t_z/T_{c0}=2.93$
in the presence of both PPB and orbital effects. b) Magnetic
Bloch wavevector $Q$ vs magnetic field. }
\label{Fig10}
\end{figure}

\begin{figure}
\caption{a) Transition line $T_c$ for $t_z/T_{c0}=3.33$
in the absence of PPB effect. b) Magnetic
Bloch wavevector $Q$ vs magnetic field. }
\label{Fig11}
\end{figure}

\begin{figure}
\caption{a) Transition line $T_c$ for $t_z/T_{c0}=3.33$
in the presence of both PPB and orbital effects. b) Magnetic
Bloch wavevector $Q$ vs magnetic field. }
\label{Fig12}
\end{figure}

\begin{figure}
\caption{a) Transition line $T_c$ for $t_z/T_{c0}=4$
in the absence of PPB effect. b) Magnetic
Bloch wavevector $Q$ vs magnetic field. }
\label{Fig13}
\end{figure}

\begin{figure}
\caption{a) Transition line $T_c$ for $t_z/T_{c0}=4$
in the presence of both PPB and orbital effects. b) Magnetic
Bloch wavevector $Q$ vs magnetic field. }
\label{Fig14}
\end{figure}

\begin{figure}
\caption{a) Transition line $T_c$ for $t_z/T_{c0}=6.67$
in the absence of PPB effect. b) Magnetic
Bloch wavevector $Q$ vs magnetic field. }
\label{Fig15}
\end{figure}

\begin{figure}
\caption{a) Transition line $T_c$ for $t_z/T_{c0}=6.67$
in the presence of both PPB and orbital effects. b) Magnetic
Bloch wavevector $Q$ vs magnetic field. }
\label{Fig16}
\end{figure}

\begin{figure}
\caption{Eigenvalue $\lambda _Q$ of the kernel $K(x,x')$ vs $Q$ for
different values of the magnetic field. The parameters are the same as
in Fig.16. (a) At low field ($H=0.3$
T), $\lambda _Q$ is independent of $Q$. At higher field, this degeneracy is
lifted in favor of $Q=0$ or $Q=G$. (b) At high enough field, $\lambda _Q$ is
maximum for either $Q=2h/v$ or $Q=G-2h/v$. }
\label{Fig17}
\end{figure}


\begin{references}
\bibitem{local} Nonetheless, the attractive interaction at the origin of the
superconductivity should be local with respect to the $z$ direction. This
condition is clearly necessary in order to obtain a reentrant
superconducting phase.
\bibitem{CB91b} More precisely, the authors of Ref.\cite{CB91} predict that
the 1D Luttinger liquid is unstable at low temperature with respect either
to a Fermi liquid (which may undergo any phase transition at lower
temperature as a normal 3D Fermi liquid) or to various kinds of ordered
ground states stabilized by coherent pair hopping.
\end{references}
\end{document}